\documentclass[reprint,superscriptaddress,amsmath,amssymb,aip]{revtex4-1}
\usepackage{tabularx}
\usepackage{graphicx}
\usepackage{dcolumn}
\usepackage{bm}
\usepackage[colorlinks=true]{hyperref}
\usepackage{siunitx}
\usepackage[dvipsnames]{xcolor}
\usepackage{microtype}
\usepackage{newtxtext,newtxmath}

\newcommand{\yield}[1]{\sigma_{\rm y}^{\rm (#1)}}

\begin{document}

\title{The role of friction in the yielding of adhesive non-Brownian suspensions}

\author{J. A. Richards}
\email{jamesrichards92@gmail.com}

\author{B. M. Guy}

\author{E. Blanco}
\affiliation{SUPA, School of Physics and Astronomy, The University of Edinburgh, Peter Guthrie Tait Road, Edinburgh, EH9 3FD, United Kingdom}

\author{M. Hermes}
\affiliation{SUPA, School of Physics and Astronomy, The University of Edinburgh, Peter Guthrie Tait Road, Edinburgh, EH9 3FD, United Kingdom}
\affiliation{Debye Institute, Utrecht University, Princetonplein 5, 3584 CC Utrecht, Netherlands}

\author{G. Poy}
\affiliation{SUPA, School of Physics and Astronomy, The University of Edinburgh, Peter Guthrie Tait Road, Edinburgh, EH9 3FD, United Kingdom}
\affiliation{Faculty of Mathematics and Physics, Jadranska 19, 1000 Ljubljana, Slovenia}

\author{W. C. K. Poon}
\affiliation{SUPA, School of Physics and Astronomy, The University of Edinburgh, Peter Guthrie Tait Road, Edinburgh, EH9 3FD, United Kingdom}

\begin{abstract}
Yielding behavior is well known in attractive colloidal suspensions. Adhesive non-Brownian suspensions, in which the interparticle bonds are due to finite-size contacts, also show yielding behavior. We use a combination of steady-state, oscillatory and shear-reversal rheology to probe the physical origins of yielding in the latter class of materials, and find that yielding is not simply a matter of breaking adhesive bonds, but involves unjamming from a shear-jammed state in which the micro-structure has adapted to the direction of the applied load. Comparison with a recent constraint-based rheology model shows the importance of friction in determining the yield stress, suggesting novel ways to tune the flow of such suspensions.
\end{abstract}


\maketitle

\section{Introduction\label{sec:intro}}

A recent paradigm shift in repulsive non-Brownian (nB) suspension rheology was inspired by the physics of jamming in dry grains. Shear thickening in such suspensions is now thought to be driven by the formation of compressive frictional contacts between neighboring particles beyond a certain critical, or onset, stress, $\sigma^*$, which overcomes the stabilizing interparticle repulsion~\cite{comtet2017pairwise,clavaud2017revealing}. Importantly, $\sigma^*$ scales roughly as the inverse square of particle size~\cite{guy2015towards}, and is readily exceeded for nB suspensions, whose flow is therefore typically dominated by frictional contacts. 
Experiments~\cite{guy2015towards,royer2016rheological} and simulations~\cite{seto2013discontinuous,mari2014shear} are captured by a phenomenological model by Wyart and Cates (WC). In the WC model a jamming volume fraction, at which the viscosity diverges, is set by a stress-dependent fraction of frictional contacts~\cite{wyart2014discontinuous}; the WC model has then been successfully extended to time-dependent flows~\cite{chacko2018dynamic,richards2019competing}.

Non-Brownian suspensions occur widely in industrial products (concrete, paint, etc.)~and their processing. The size of nB particles ($\gtrsim \SI{10}{\micro\meter}$) means that residual van der Waals attraction is all but inevitable despite steric or charge stabilization~\cite{guy2015towards}. If strong enough, such interaction gives rise to a yield stress, $\sigma_{\rm y}$, below which suspensions cannot flow~\cite{brown2010generality}. Such behavior occurs in, e.g., mine tailings and mineral slurries~\cite{zhou1995yield,nguyen1998application} or molten chocolate~\cite{blanco2019conching}. Controlling $\sigma_{\rm y}$ in suspensions is important for their stability during transport and shaping, and for suspending various macroscopic particulates such as sand~\cite{ovarlez2015flows}. 

We have recently extended the WC framework to model the flow of nB suspensions with more varied particle-level interactions and hence to describe suspensions with a finite yield stress~\cite{guy2018constraint}. We treat interparticle friction as a constraint to relative sliding between particles that switches on with increasing stress. Adding a second constraint restricting interparticle rotation that is removed with increasing stress enables us to predict all classes of flow curves observed in the literature. A paradigmatic example of the second kind of constraint is adhesion: `sticky' finite-area contacts constraining interparticle rotation that can be broken if the applied stress exerts a critical torque on neighboring particles~\cite{heim1999adhesion,pantina2005elasticity}.  

This `constraint rheology' of nB suspensions has a number of non-trivial implications, which are either discussed cursorily or remain implicit in our previous work~\cite{guy2018constraint}. Here, we present an extensive rheological study of a model adhesive nB suspension, cornstarch in oil, to highlight and discuss one such implication, that yielding in adhesive nB suspensions should be qualitatively different from corresponding phenomena in Brownian (or colloidal) suspensions. In the latter, friction typically plays no role and the attraction between particles, which is described by a potential, does not by itself constrain interparticle rotation. There are similarities in the yielding phenomenology of the two kinds of systems. For example, we find that an adhesive nB suspension yields in two steps under certain rheological protocols, recalling attractive colloidal glasses~\cite{pham2006yielding}. However, such resemblance turns out to be superficial, and hides a profound difference in the underlying physical mechanisms. 

Specifically, interparticle friction plays a key role in the genesis of a yield stress and in determining its magnitude. This role is unobvious in the form of the steady-state flow curve. However, the dependence of the steady-state yield stress, $\yield{ss}$, on the solid volume fraction, $\phi$, points to a role for friction: $\yield{ss}(\phi)$ diverges before random close packing, $\phi_{\rm rcp}$, at a lower frictional jamming point, $\phi_{\mu}$. In contrast, large-amplitude oscillatory rheology returns a substantially lower yield stress, $\yield{os}$, which does diverge at $\phi_{\rm rcp}$. Finally, shear reversal experiments reveal that upon changing the direction of shear there is a transient yielding event at an intermediate `transient reversal yield stress', $\yield{tr} < \yield{ss}$. Interestingly, $\yield{tr}(\phi)$ follows the oscillatory yield stress at low $\phi$, before increasing to approach the steady-state yield stress at higher $\phi$.

Thus, the yield stress in adhesive nB suspensions is protocol dependent~\cite{nguyen2006yield}, which we connect with two fundamental features of nB suspensions. First, without thermal motion, sticky particles not already in contact will {\it not} encounter each other to build higher-order stress-bearing structures except under external deformation. So, secondly, given the low $\sigma^*$ for nB particles, such deformation will almost always involve stress $> \sigma^*$, and so will bring interparticle friction into play. Therefore, adhesion seldom acts alone in nB suspensions. Using steady-state, oscillatory and reversal rheology together allows us to illustrate these two features and highlight the differences between adhesive nB suspensions and colloidal suspensions with interparticle potential attraction. 

\section{Constraint rheology\label{sec:constraint}}

For later use, we first recast constraint rheology~\cite{guy2018constraint} for the specific case of a nB suspension with friction (constraining sliding) and adhesion (constraining rolling). Stripped to its bare essentials, WC proposes that increasing stress progressively makes frictional contacts, which pose additional constraints on interparticle motion by removing the freedom of contacting particles to slide past each other. In turn this lowers the jamming volume fraction, causing shear thickening. Following WC’s constraint-motivated train of thought, Guy et al.~\cite{guy2018constraint} propose that adhesion removes the freedom of contacting particles to roll on each other, thus lowering the jamming volume fraction; however, increasing stress progressively removes such constraints, thereby allowing the jamming volume fraction to rise, leading to shear thinning or even yielding from a shear-jammed state.

Specifically, following WC, the fraction of frictional contacts increases with stress, $\sigma$, according to
\begin{equation}
f(\sigma) = \exp\left[ -\left(\frac{\sigma^*}{\sigma} \right)^\beta \right], \label{eq:M1}
\end{equation}
with $\beta$ an exponent describing how rapidly $f$ increases from 0 ($\sigma \ll \sigma^*$) to 1 ($\sigma \gg \sigma^*$). Additionally, stress decreases the fraction of adhesive constraints according to
\begin{equation}
a(\sigma) = 1 -  \exp\left[ -\left(\frac{\sigma_a}{\sigma} \right)^\kappa \right], \label{eq:M2}
\end{equation}
with $\sigma_a$ setting the stress scale for breaking adhesive contacts and $\kappa$ another exponent that captures how rapidly adhesive contacts are broken and the suspension shear thins. The jamming volume fraction is a function of these two variables, $\phi_{\rm J} = \phi_{\rm J}(f,a)$. 

This function is well known in two limits. The maximum amorphous packing for frictionless, adhesionless hard spheres is random close packing, $\phi_{\rm J}(f\!=\!0,~ a\!=\!0) \equiv \phi_{\rm rcp} \approx 0.64$; the corresponding quantity for frictional hard spheres is $\phi_{\rm J}(f\!=\!1,~ a\!=\!0) \equiv \phi_{\mu} \approx 0.55$ for highly frictional particles~\cite{jerkins2008onset, silbert2010jamming}. The limits with all-adhesive contacts are less well known. Simulating ballistic deposition~\cite{liu2015adhesive,liu2017equation} finds `adhesive loose packing' at $\phi_{\rm J}({f\!= 1},~{ a\!=\!1}) \equiv \phi_{\rm alp} \approx 0.15$ and `adhesive close packing' at $\phi_{\rm J}({f \!= \!0},~{ a\! =\! 1}) \equiv \phi_{\rm acp} \approx 0.51 = \phi_{\mu}$ given current levels of uncertainties, including likely protocol dependence. 

Following WC, we interpolate to give 
\begin{eqnarray*}
\phi_{\rm J}(f,a) = af \phi_{\rm alp} + a(1-f) \phi_{\rm acp} &+& (1-a)f \phi_{\mu} \\
&+& (1-a)(1-f) \phi_{\rm rcp}.
\end{eqnarray*}
For us, $\sigma^* \to 0$ and $f = 1$ always, so that 
\begin{equation}
\phi_{\rm J}(f = 1, a) = a \phi_{\rm alp} + (1-a)\phi_{\mu}. \label{eq:M3}
\end{equation}
Finally, as in WC, we take the relative viscosity to be
\begin{equation}
\eta_{\rm r} = \left[ 1 - \frac{\phi}{\phi_{\rm J}(a,f)} \right]^{-2}. \label{eq:M4}
\end{equation}
Together, Eqs.~\ref{eq:M1}, \ref{eq:M3} and \ref{eq:M4} describe suspensions that yield and shear thin, class 1 flow curves in the terminology of Ref.~\cite{guy2018constraint}. 

To conclude this section, we make a proposal for terminology. In a canonical colloidal suspension, particles do not contact. Their interaction, described as the gradient of a potential, does {\it not} constrain interparticle rolling and suspensions of such particles are termed `cohesive'~\cite{singh2019yielding}; this is distinct from the sticky contacts that concern us in this work. We propose to mark this important difference by strictly distinguishing between the terms `attraction' and `adhesion', with the latter denoting finite contact area with a concomitant rolling constraint. We find that markedly different physics underlies the yielding of adhesive nB suspensions and attractive suspensions~\cite{pednekar2017simulation,singh2019yielding}.

\section{Experimental system and methods\label{sec:methods}}

Cornstarch in aqueous media is a model for the rheology of purely repulsive nB suspensions, showing characteristic friction-driven shear thickening at a fixed onset stress~\cite{richards2019competing,hermes2016unsteady}. When cornstarch is dispersed in non-aqueous solvents, shear thickening is no longer observed and a finite yield stress arises~\cite{freundlich1938dilatancy,galvez2017dramatic}. Adhesive particle interactions could originate from Van der Waals forces, hydrogen bonding~\cite{james2019tuning} or even capillary forces~\cite{koos2011capillary} from adsorbed water in cornstarch particles~\cite{han2017measuring}. We disperse previously-employed~\cite{richards2019competing,hermes2016unsteady} cornstarch (Sigma Aldrich) in sunflower oil (Flora) to form a model adhesive nB suspension. The particles have diameter $d \approx \SI{14}{\micro \meter}$ and polydispersity $\approx 40\%$ (from static light scattering~\cite{hermes2016unsteady}) and density of $\rho_{\rm p} = \SI{1.45}{\gram \centi \meter ^{-3}}$. The sunflower oil has viscosity and density $\eta_{\rm f} = \SI{62}{\milli \pascal \second}$ and $\rho_{\rm f} = \SI{0.92}{\gram \centi \meter^{-3}}$ respectively at \SI{20}{\celsius}. Cornstarch was dispersed by vortex mixing and stirring until visually homogeneous before roller mixing for $\gtrsim \SI{2}{\hour}$. Cornstarch does not swell in non-aqueous solvents~\cite{chen2019discontinuous}, so that swelling corrections~\cite{han2017measuring} are unnecessary. 

We used a strain-controlled ARES-G2 rheometer for steady-state and oscillatory measurements, and a controlled-torque DHR-2 rheometer for shear reversal experiments (both TA Instruments). Measurement geometry selection requires care. The \SI{22}{\micro\meter} truncation gap of our cone-plate geometry was too small for our particles. In a Couette cell, sedimentation can give rise to an apparent yield stress~\cite{fall2009yield}, while stress variation across the gap can lead to spatial inhomogeneities~\cite{fall2010shear}. We therefore used parallel plates (radius $R=\SI{20}{\milli \meter}$, gap height $h=\SI{1}{\milli \meter}$) with cross-hatching (\SI{0.25}{\milli \metre} serrations) to reduce slip. 

The steady-state rheology of our samples can be probed within a `window' of shear stresses and rates, Fig.~\ref{fig:ss} (white region). The maximum stress in this window is set by sample fracture, which occurs for us at $\sigma_{\max} \approx \SI{180}{\pascal}$~\cite{guy2015towards}. The low shear rate limit, $\dot\gamma_{\rm min}$, is set by the experimental time, which is limited by, e.g., drying or sedimentation of the sample, the latter setting a minimum stress at $\sigma_{\rm min} = (\rho_{\rm p} -\rho_{\rm f}) g d \approx \SI{0.05}{\pascal}$, which is larger than the $\sigma_{\rm min} \approx \SI{0.01}{\pascal}$ set by the torque resolution of ARES-G2. Finally, the maximum shear rate, $\dot\gamma_{\rm max}$, is set by inertial sample rejection. 

\section{The steady-state yield stress\label{sec:steady}}

\begin{figure}
    \centering
    \includegraphics{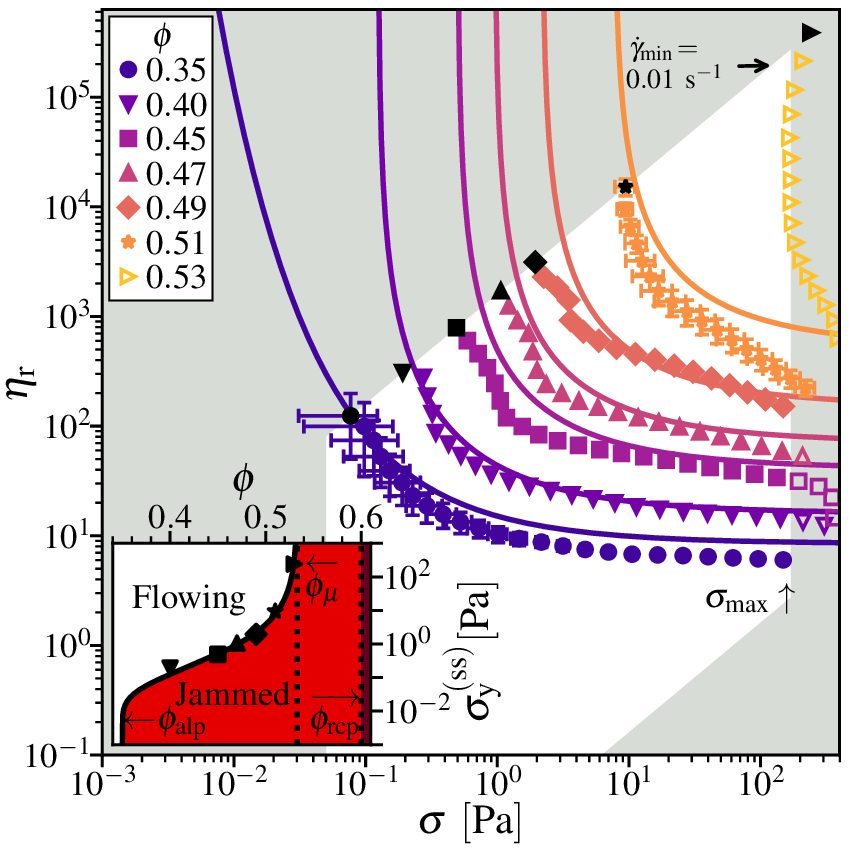}
    \caption{Cornstarch-in-oil flow curves under imposed shear rate, $\dot{\gamma}$: relative viscosity, $\eta_{\rm r} =\eta/\eta_{\rm f}$, vs.~shear stress, $\sigma$. White shading, observable `window' (see text for details). See legend for volume fractions, $\phi$. Black points, $\sigma(\dot{\gamma}_{\min}) \equiv \yield{ss}$; open symbols, unreliable measurements outside `window'. Flow curves are averaged over 3 runs for $\phi \leq 0.47$ and 2 for $\phi=0.51$ to estimate reliability, standard deviation shown when larger than marker. Lines, fit to constraint-based rheology model~\cite{guy2018constraint}, Eqs.~(\ref{eq:M2})-(\ref{eq:M4}), with $\phi_{\mu} = 0.533$, $\phi_{\rm alp}=0.35$, $\sigma_a = \SI{0.2}{\pascal}$ and $\kappa =0.55$. Inset: steady-state yield stress, $\yield{ss}$, as a function of $\phi$. Points, $\yield{ss}$, note that $\phi=0.35$ is excluded as no yield stress is measured under imposed stress; solid line, yield stress from constraint-based model setting $\phi_{\rm J}=\phi$. Shading: red, jammed at steady state, $\phi_{\rm J} < \phi$; white, flowing at steady state, $\phi_{\rm J} > \phi$; and maroon, $\phi > \phi_{\rm rcp} \sim 0.6$, dispersion not possible.}
    \label{fig:ss}
\end{figure}

To measure flow curves, we presheared suspensions of various $\phi$ at $\dot{\gamma}=\SI{10}{\second^{-1}}$ or $\dot{\gamma}(\sigma_{\max})$, whichever is lower, and then dropped the imposed rate to $\dot{\gamma}_{\min}$ to begin an up-sweep at 6 points per decade with a time interval of either \SI{10}{\second} or a longer interval to accumulate a strain of $\gamma = 10$. In a parallel-plate geometry, the imposed angular velocity, $\Omega$, and the measured torque, $M$, give the rim shear rate, $\dot{\gamma}=\Omega R / h$, and the corrected stress, $\sigma=(M / 2 \pi R^3)(3 + \mathrm{d} \ln M/\mathrm{d} \ln \Omega)$~\cite{macosko1994rheology}. The relative viscosity, $\eta_{\rm r} = \sigma/\dot{\gamma} \eta_{\rm f}$, as a function of $\sigma$ at different $\phi$, Fig.~\ref{fig:ss}, shows significant shear thinning at $\phi \gtrsim 0.35$: $\eta_{\rm r}$ decreases with $\sigma$ to approach what appears to be a high-shear plateau, which, however, is obscured at higher $\phi$ by sample fracture at $\sigma_{\max} \approx \SI{180}{\pascal}$. Similar flow curves have been widely reported in nB suspensions, including various 2-\SI{25}{\micro\meter} refractory particles and cocoa powder \cite{zhou1995yield}, 3-\SI{5}{\micro\meter} PMMA \cite{heymann2002solid} and molten chocolate ($\approx \SI{15}{\micro\meter}$ sugar crystals suspended in a triglyceride) \cite{blanco2019conching}.

Such flow curves are typically taken to evidence interparticle attraction, whose strength is estimated by plotting  $\sigma(\dot\gamma)$ and extrapolating to $\dot\gamma = \SI{0}{\per\second}$ using an empirical model, e.g.~Herschel-Bulkely or Casson, to obtain the steady-state yield stress, $\yield{ss}$. We estimate $\yield{ss}$ as the stress at the lowest accessed shear rate, i.e.,~$\yield{ss} = \sigma(\dot\gamma_{\min} = \SI{0.01}{\per \second})$. For $\phi = 0.35$, this produced a finite $\yield{ss}$, but tests under controlled stress found that the sampled flowed at all applied stresses ($\sigma > \SI{3}{\milli \pascal}$), so that in fact for $\phi = 0.35$ we take $\yield{ss} = \SI{0}{\pascal}$.

The role of friction in the yielding of our suspensions is revealed by the $\phi$ dependence of  $\yield{ss}$, Fig.~\ref{fig:ss}~(inset), which appears to diverge at $\phi \approx 0.54$. The frictionless and frictional jamming points of cornstarch in aqueous solvents are $\phi_{\rm rcp} \approx 0.60$~\cite{han2017measuring} and $\phi_{\mu} \approx 0.50$ respectively. The latter is estimated from multiplying $\phi_{\mu} \approx 0.6$ for non-aqueous cornstarch by the measured weight-fraction ratio of 0.84 for random loose to random close packing for aqueous cornstarch~\cite{richards2019competing}. We therefore take the $\phi \approx 0.54$ at which $\yield{ss}(\phi) \to \infty$ to be the {\it frictional} jamming point, $\phi_{\mu}$, of cornstarch in oil. Consistent with this, we could make samples at $\phi > 0.54$; but these samples showed unsteady stick-slip flow or fracture, recalling similar behavior above $\phi_{\mu}$ in aqueous cornstarch suspensions~\cite{hermes2016unsteady}. 

If our proposal that $\yield{ss}\to\infty$ at the frictional jamming point is correct, then the absence of shear thickening in our flow curves implies that our suspensions at $\phi < \phi_{\mu}$ always flow with frictional contacts after yielding ($f = 1$ because $\sigma^* \to 0$). Indeed, we find that Eqs.~\ref{eq:M2}-\ref{eq:M4} with $\phi_{\mu} = 0.533, \phi_{\rm alp} = 0.35, \sigma_a = \SI{0.2}{\pascal}$ and $\kappa = 0.55$ can credibly account for both our flow curves [Fig.~\ref{fig:ss} (lines)] and the $\yield{ss}(\phi)$ inferred from them (inset). Yielding in this suspension is then a matter of overcoming shear jamming due to a combination of adhesion and friction in the interval $\phi_{\rm alp} < \phi < \phi_{\mu}$. This contrasts with colloidal systems, where a gelled state can form without external perturbation due to thermal motion, whereas in a non-Brownian system structure can only form under external mechanical perturbation such as shear. Unjamming this shear-jammed state requires breaking adhesive bonds with stress, removing constraints progressively until $\phi_{\rm J}$ exceeds the sample $\phi$ at some critical stress that we identify as $\yield{ss}$. This yield stress first arises at the minimum jamming volume fraction, $\phi_{\rm J}(a\!=\!1,~f\!=\!1) = \phi_{\rm alp}$ and diverges at the maximum value of $\phi_{\rm J} = \phi_{\mu}$.  

Our $\phi_{\rm alp} = 0.35$ is considerably higher than the $\approx 0.15$ found in recent simulations \cite{liu2015adhesive,liu2017equation}. This may partly reflect particle properties (monodisperse spheres vs.~polydisperse cornstarch grains), but it may also reflect fundamental physics. If the solid-like state at $\sigma < \yield{ss}$ is due to jamming, then its properties could depend on how jamming was induced in the first place. As ballistic deposition~\cite{liu2015adhesive,liu2017equation} and steady shear give different jammed states, the rigidity percolation threshold, $\phi_{\rm alp}$, may also differ.

Since our particle contacts are sticky, frictional and of finite area, they form rigid bonds, so that contact and rigidity percolation coincide, and the value of $\phi_{\rm alp}$ should represent this coincident percolation threshold under shear. Indeed, our $\phi_{\rm alp}$ is similar to the percolation threshold found for frictionless and adhesionless nB suspensions (of monodisperse spheres) under shear~\cite{gallier2015percolation}, which furthermore is not strongly affected by the presence of additional constraints such as friction~\cite{gallierThesis}. In contrast, for attractive interactions without bond rigidity, percolation and rigidity percolation differ~\cite{zhang2019correlated}.

\section{The oscillatory yield stress\label{sec:oscillatory}}

\begin{figure}
    \centering
    \includegraphics{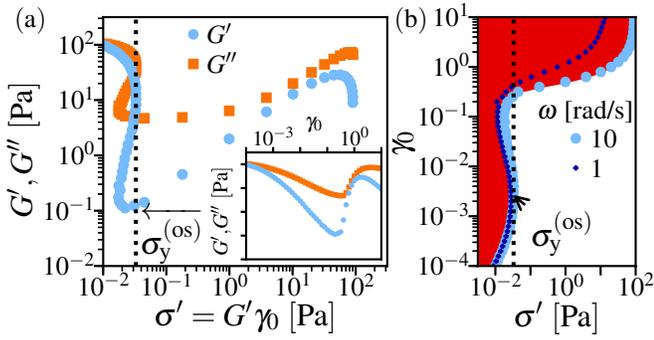}
    \caption{Oscillatory rheology at $\phi=0.51$, for a decreasing imposed strain amplitude, $\gamma_0$. (a)~Elastic modulus, $G^{\prime}$ (light blue circle), and loss modulus, $G^{\prime\prime}$ (orange square), as a function of elastic stress, $\sigma^{\prime} = G^{\prime}\gamma_0$. Data is taken at 10 points per decade from $\gamma_0 = 10$ to $\gamma_0 = 10^{-4}$ at an (angular) frequency $\omega = \SI{10}{\radian \per \second}$ with 1 delay cycle and 7 measurement cycles. Inset: moduli~vs.~$\gamma_0$. (b)~Load curve: strain amplitude, $\gamma_0$, vs.~elastic stress, $\sigma^{\prime}$, at $\omega=\SI{1}{\radian\second^{-1}}$ (dark blue) taken with 1 delay cycle and 1 measurement cycle and $\SI{10}{\radian\second^{-1}}$ (light blue), as in (a). The oscillatory yield stress, $\yield{os}$ (dotted line), is identified from the value of $\sigma^{\prime}$ where ${\rm d}\sigma^{\prime}/{\rm d}\gamma_0 = 0$ with the minimum $\gamma_0$ at $\omega = \SI{10}{\radian \second^{-1}}$. The oscillatory yield stress is also indicated in (a) for comparison. Red, predicted jammed region, and white, predicted (transient) flow region.}
    \label{fig:osc}
\end{figure}
                                                             
To disentangle the entwined roles of friction and adhesion in steady-state flow, we performed oscillatory rheology. Applying sinusoidal shear at an (angular) frequency of $\omega = \SI{10}{\radian \per \second}$, we measured the storage and loss moduli, $G^{\prime}$ and $G^{\prime\prime}$, in a downsweep of strain amplitude, $\gamma_0$, starting from either $\gamma_0 = 10$, or the highest strain amplitude reachable without fracture; this removed loading effects and ensured repeatability. The measured $G^\prime(\gamma_0)$ and $G^{\prime\prime}(\gamma_0)$ at $\phi = 0.51$, Fig.~\ref{fig:osc}(a), show a slow decrease with $\gamma_0$ but no sudden yielding. The main figure replots this data against the so-called `elastic stress'~\cite{yang1986some}, $\sigma^\prime = G^\prime \gamma_0$. If $G^{\prime}$ was frequency independent and $G^{\prime\prime}$ scaled viscously ($\propto \omega$), $\sigma^\prime$ would represent the stress at zero frequency. In this representation, it is clear that the sample yields -- the moduli drop abruptly -- at some critical stress. This is confirmed by plotting $\gamma_0(\sigma^\prime)$, an approximate static (i.e.~$\dot{\gamma}$ or $\omega \to 0$) stress-strain curve, Fig.~\ref{fig:osc}(b). This function is $\omega$-independent up to an oscillatory yield stress, $\yield{os} \approx \SI{0.03}{\pascal}$, where $\gamma_0$ makes an abrupt jump by more than three orders of magnitude over a very small interval of $\sigma^\prime$.

In the raw data we find $G^{\prime}$ is smaller than $G^{\prime\prime}$ for all $\gamma_0$ at $\omega = \SI{10}{\radian \per \second}$. This is simply because $G^{\prime\prime} \sim \eta \omega$ while $G^{\prime}$ is essentially $\eta$-independent, and we have used a high-viscosity solvent to bring the relevant phenomena into the stress window of our rheometer.  At the lower frequency of $\omega = \SI{1}{\radian \per \second}$ we indeed recover $G^{\prime} > G^{\prime\prime}$ at small strains, as expected for a `solid-like' sample.

\begin{figure}
    \centering
    \includegraphics{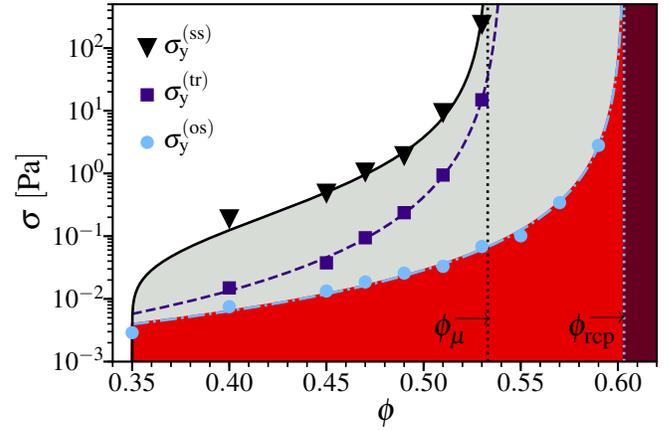}
   \caption{$\phi$ dependence of yield stresses. Symbols: $\blacktriangledown$, steady-state yield stress, $\yield{ss}$; $\blacksquare$, transient yield stress upon shear reversal, $\yield{tr}$; and $\bullet$, oscillatory yield stress, $\yield{os}$. Lines: solid black, yield stress of constraint-based model, parameters as in Fig.~\ref{fig:ss}; dashed purple, fit of $\yield{tr}$ to $A (1-\phi/\phi_{\rm crit})^{-l}$, with $\phi_{\rm crit} = 0.541 \pm 0.002$ ($A = 3 \times 10^{-4}$ and $l = 2.8$); dot-dashed light blue, fit of $\yield{os}$ to $A (1-\phi/\phi_{\rm rcp})^{-l}$ to extract $\phi_{\rm rcp}=0.603\pm 0.003$ ($A=5 \times 10^{-4}$ and $l=2.2$); black dotted, $\phi_{\mu}=0.533$ from constraint-based model; and light blue dotted, $\phi_{\rm rcp}$ from $\yield{os}$ divergence. Shaded regions: white, continuous flow; gray, jammed at steady-state but transient flow possible ($\yield{os}(\phi) < \sigma < \yield{ss}(\phi)$) indicating range of possible protocol-dependent yield-stress measurements; red, jammed at steady-state and no transient flow; and maroon, no dispersion possible ($\phi > \phi_{\rm rcp}$).}
    \label{fig:comp}
\end{figure}

The dependence of the oscillatory yield stress on suspension concentration, $\yield{os}(\phi)$, is shown in Fig.~\ref{fig:comp}, where we have also replotted the corresponding function for the steady-state yield stress, $\yield{ss}(\phi)$, for comparison. Two features immediately stand out. First, $\yield{os} \ll \yield{ss}$, by one and a half orders of magnitude at $\phi = 0.4$ and rising to four orders of magnitude at $\phi \lesssim 0.53$. Secondly, we can measure a finite $\yield{os}$ considerably beyond $\phi_{\mu} \approx 0.53$. Indeed, fitting to $A(1-\phi/\phi_c)^{-l}$ shows that $\yield{os}(\phi)$ diverges (with $A = 5 \times 10^{-4}$ and $l = 2.2$) at $\phi_c \approx 0.603$, which is $\phi_{\rm rcp}$ for aqueous cornstarch.

These features suggest that oscillatory shear removes frictional contacts to enable probing of yielding to a frictionless state, which therefore does not jam until $\phi_{\rm rcp}$. The oscillatory shear applied is sufficient to break and mobilize adhesive bonds. What we observe can therefore be usefully compared with the way repeated oscillatory shear removes or relaxes contacts in non-adhesive nB systems~\cite{corte2008random,ness2018shaken}, and with the shaking dry grains to compactify the packing from (frictional) random loose packing to (frictionless) random close packing~\cite{baker2010maximum}. The magnitude of  $\yield{os}$ therefore reflects the adhesive contact strength alone without the effects of friction. However, $\yield{os}$ does not directly measure the adhesive bond strength, $\sigma_a = \SI{0.2}{\pascal}$, that we have previously extracted from the steady-state flow curves, because $\yield{os}$ is a collective property reflecting $\sigma_a$ and suspension structure. 

After the abrupt rise in $\gamma_0(\sigma^\prime)$ at $\yield{os}$, Fig.~\ref{fig:osc}(b), the function bends over at $\gamma_0 \approx 0.5$, whereupon $\sigma^{\prime}$ rises rapidly with $\gamma_0$. This suggests that the system rejams with strain after yielding at $\yield{os}$, so that yielding at $\yield{os}$ is only transient. We attribute the rejamming to the remaking of frictional contacts, which occurs at a strain where frictional contacts have been found to reform after reversal in shear-thickening nB suspensions at a similar volume fraction relative to $\phi_{\rm rcp}$~\cite{lin2015hydrodynamic}. However, because we impose a finite-amplitude sinusoidal strain, we cannot access true jamming ($\dot\gamma = 0$), and can only infer it in a manner analogous to inferring jamming in shear-thickening nB suspensions under imposed shear rate~\cite{mari2014shear}. Moreover, our data at $\sigma > \yield{os}$ is strongly frequency dependent, Fig.~\ref{fig:osc}(b), with $\sigma^\prime$ shifting linearly with $\omega$, indicating viscous behavior. Thus, the second upturn in $\gamma_0(\sigma^{\prime})$ at large $\sigma^{\prime}$ should \emph{not} be directly interpreted as a second yielding event. 

\begin{figure}
    \centering
    \includegraphics{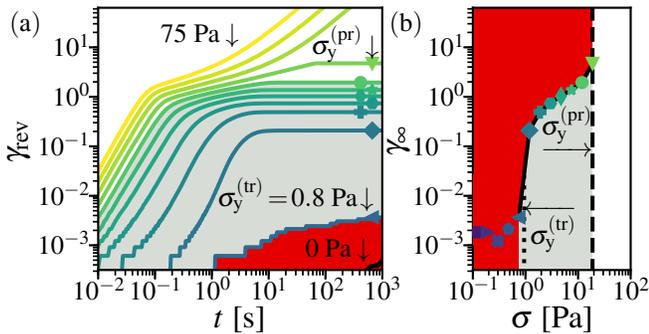}
\caption{Shear reversal at $\phi=0.51$. (a)~Time-dependent strain response after application of stress in the reverse direction, $\gamma_{\rm rev}(t)$, for imposed stresses, $\sigma$, spaced logarithmically at 5 points per decade from \SI{0.075}{\pascal} (data shown from \SI{0.8}{\pascal}) to \SI{75}{\pascal} (teal to yellow) and \SI{0}{\pascal} (black). Stress given by color in (b) for states that are jammed at $t= \SI{1000}{\second}$; for flowing states $\sigma =$ 30, 47 and \SI{75}{\pascal}. Shading: red, range of responses showing creep ($\sigma < \yield{tr}$); gray, transient yielding response ($\yield{tr} < \sigma < \yield{pr}$); and white, permanently flowing ($\sigma > \yield{pr}$). (b)~Long-time limiting strain, $\gamma_\infty(\sigma) \equiv \gamma_{\rm rev}(t=\SI{D3}{\second})$, vs.~$\sigma$ (symbols). Dotted line, $\yield{tr}$, indicates the stress for transient yielding, identified from the largest increase in $\log(\gamma_\infty)$ with $\log(\sigma)$; dashed line indicates, $\yield{pr}$, the stress to permanently flow. Shading: white, flowing states; gray, transiently flowing states (will jam at $\gamma_\infty$); and, red, inaccessible jammed states.}
\label{fig:rev}
\end{figure}

\section{The transient reversal yield stress\label{sec:reversal}}

A protocol that does show transient yielding, re-jamming, and a second, permanent yielding is shear reversal. It has been used to reveal the role of friction in repulsive nB suspensions~\cite{lin2015hydrodynamic,peters2016rheology}: abruptly reversing the shear direction during steady-state shear breaks all frictional contacts, which are only reformed when a reversed strain of order unity has been accumulated. 

To reach a well-defined initial state, samples were presheared just below the fracture stress, $\sigma_{\max}$, for \SI{100}{\second}, left quiescent for \SI{100}{\second}, and then stressed at a constant $\sigma < \sigma_{\max}$ for \SI{1000}{\second} in the opposite direction. We work in terms of the rim strain, $\gamma$, and apparent stress, $\sigma=3 M /2\pi R^3$, correct at yielding where $\mathrm{d} \ln M / \mathrm{d} \ln \Omega = 0$. The time-dependent strain response in the new direction, $\gamma_{\rm rev}(t)$, was measured over a range of stresses. Data for $\phi=0.51$, Fig.~\ref{fig:rev}(a), are typical.

At $\sigma < \SI{0.8}{\pascal}$, we find a sub-linear growth of $\gamma_{\rm rev}(t)$, or creep; its occurrence at $\sigma = \SI{0}{\pascal}$ shows that creep is a remnant of preshear. Above the transient reversal yield stress $\yield{tr}=\SI{0.8}{\pascal}$, the suspension transiently unjams and flows at constant acceleration, $\gamma_{\rm rev} \propto t^2$, which reflects instrument inertia. Below $\sigma = \SI{20}{\pascal}$, the flowing suspension then re-jams at $\gamma_{\rm rev} \approx {O}(1)$. This strain is not recoverable. Above a permanent reversal yield stress $\yield{pr} = \SI{20}{\pascal}$, the suspension unjams again, now permanently yielding to continuous, viscous flow with $\gamma_{\rm rev} \propto t$.

\begin{figure}[t]
    \centering
    \includegraphics{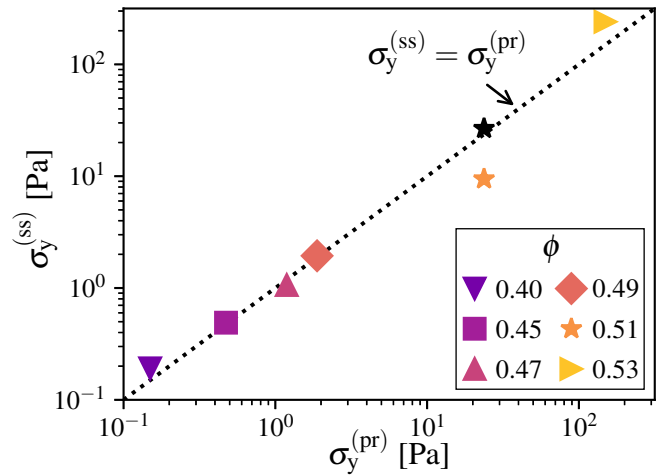}
\caption{Comparison of the steady-state yield stress, $\yield{ss}$ (colored), and the permanent reversal yield stress, $\yield{pr}$ at different volume fractions (see legend). The dashed line has unit slope. For $\phi=0.51$ a measurement of the peak stress measured at $\dot{\gamma}_{\min}$ after preshear is also shown (black star); these measurements do not differ at other $\phi$.}
\label{fig:yieldcomp}
\end{figure}

Figure~\ref{fig:rev}(b) shows the long-time limiting strain, $\gamma_\infty$, for $\sigma < \SI{20}{\pascal}$ as a function of $\sigma$. (Continuous flow at higher $\sigma$ means $\gamma_\infty \to \infty$.) The two-stepped form of $\gamma_\infty(\sigma)$ recalls $\gamma_0(\sigma^\prime)$ measured using the oscillatory protocol, Fig.~\ref{fig:osc}(b). Now, however, all states in $\gamma_{\infty}$ are jammed, with well defined plateaus in $\gamma_{\rm rev}(t)$ up to $\gamma_{\infty} \approx \mathcal{O}(1)$, so that the second upturn in $\gamma_\infty(\sigma)$ at $\yield{pr}$ indeed evidences a second yield stress. We compare the value of this yield stress with the steady-state yield stress, $\yield{ss}$, at a range of concentrations in Fig.~\ref{fig:yieldcomp}. Across $\phi$,  they are comparable to within experimental uncertainties, Fig.~\ref{fig:yieldcomp}. 

An exception occurs at $\phi=0.51$, where $\yield{pr}$ corresponds more closely to the peak stress measured at $\dot{\gamma}_{\min}$ after preshear, Fig.~\ref{fig:yieldcomp}~ (black symbol). This peak may be indicative of banding~\cite{divoux2011stress}, suggesting that the suspension may yield into a banded state above $\yield{pr}$. This single discrepancy does not impact our conclusion that $\yield{ss}$ and $\yield{pr}$ are comparable,  and both diverge at $\phi_{\mu}$. This agreement is unsurprising: above $\yield{pr}$, the system flows continuously under the reversed applied stress and a frictional contact network is fully formed ($f=1$), so that $\yield{pr}$ reflects both friction and adhesion. 

More interestingly, the dependence on concentration of the lower, transient reversal yield stress, $\yield{tr}(\phi)$, is intermediate between that of $\yield{os}(\phi)$ and $\yield{ss}(\phi)$. Fitting to $A(1-\phi/\phi_c)^{-l}$ gives a curve (with $A = 3 \times 10^{-4}$ and $l = 2.8$) that approaches $\yield{os}(\phi)$ from above at low $\phi$, and approaches $\yield{ss}(\phi)$ from below at high $\phi$, diverging at $\phi_c = 0.541 \pm 0.002$, which is the same as $\phi_{\mu}$ inferred from $\yield{ss}(\phi)$ to within experimental uncertainties. 

\begin{figure}
    \centering
    \includegraphics{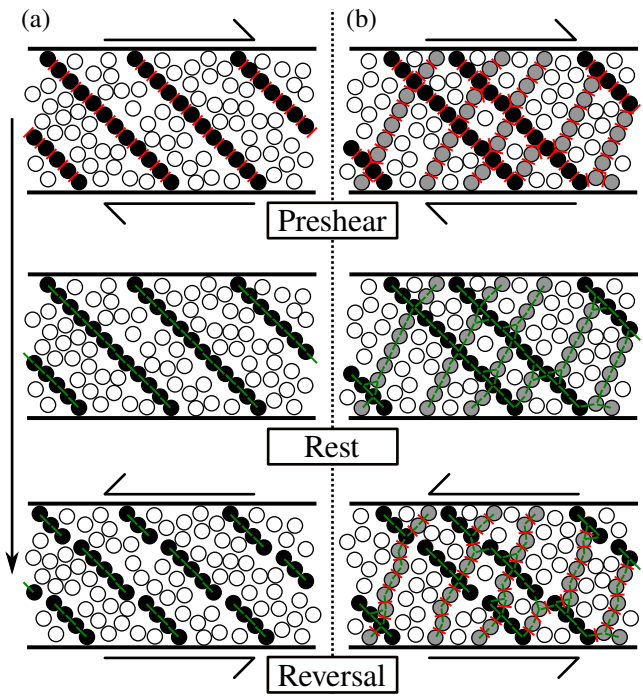}
\caption{Schematic illustration of fragility in adhesive nB suspensions under shear reversal (after Cates et al.~\cite{cates1998jamming}). (a)~Microstructure and particle interactions for moderate volume fractions, $\phi_{\rm alp} < \phi \ll \phi_{\mu}$, during: high-stress preshear (top), rest (middle) and shear reversal (bottom). Half-headed arrows indicate the direction of the applied shear stress, vertical arrow order of steps in protocol. Particles: filled black, in `force-chains' during preshear; white filled, spectator particles. Red lines show compressive frictional contacts, and green lines show adhesive contacts. (b)~Corresponding microstructure and particle interactions at a high volume fraction $\phi \lesssim \phi_{\mu}$. Filled gray particles in supporting contact network during preshear.}
\label{fig:fragile}
\end{figure}

The behavior of $\yield{tr}$ can be understood using ideas originally invoked to explain the `fragility' of shear-jammed states in repulsive nB suspensions \cite{cates1998jamming}, Fig.~\ref{fig:fragile}. Consider first the suspension structure during preshear. The high applied stress ($\sigma \gg \sigma_a$) breaks all adhesive contacts, and the microstructure resembles that of a purely frictional nB suspension at the same $\phi$, where at a high enough $\phi$, sample-spanning force-chains develop in the compressive direction [filled black circles in Fig.~\ref{fig:fragile}(a)] in which particles form frictional contacts [highlighted in red in Fig.~\ref{fig:fragile}(a)]~\cite{mari2014shear}. Simulations of hard spheres find such contact percolation at $0.3 \lesssim \phi \lesssim 0.4$\cite{gallier2015percolation}, irrespective of the presence or absence of friction \cite{gallierThesis}. These force chains, however, do not lead to shear jamming, because as they buckle under applied stress the suspension is not dense enough for other, stabilizing frictional contacts to form. 

Such reinforcement becomes possible as $\phi \to \phi_{\mu}$, giving rise to `supporting' frictional force chains [filled gray circles in Fig.~\ref{fig:fragile}(b)]~\cite{mari2014shear}, which leads to shear-jamming at $\phi_{\mu}$ in purely frictional nB suspensions~\cite{seto2019shear}. In a frictional nB suspension, the compressional force chains are broken upon reversal, and the supporting force chains are not well-aligned enough to the new compression direction to cause shear jamming immediately upon reversal. Frictional nB suspensions are therefore `fragile' -- they are only jammed (solid like) relative to a particular driving stress component.

Upon the cessation of preshear in an adhesive suspension, all the contacts become adhesive as $\sigma \ll \sigma_a$ [highlighted in green in Fig.~\ref{fig:fragile}(c) and (d)]. Consider what happens when shear is applied in the reverse direction. Now, even at $\phi$ significantly below $\phi_{\mu}$, system-spanning adhesive chains exist that can bear finite tensile stress. These are the sole load-bearing structures at low volume fractions, Fig.~\ref{fig:fragile}(e). Yielding at $\yield{tr}$ therefore involves breaking tensile, rather than compressive, contacts, so that friction is not important, and $\yield{tr} \approx \yield{os}$. As $\phi \to \phi_{\mu}$, however, frictional force chains come into being immediately upon reversal, these being originally created as `supporting' frictional force chains during preshear. Although not entirely aligned with the new compressive direction, the presence of adhesion can stabilize them to a near-compressive load, similar to the argument for particles of finite softness~\cite{cates1998jamming}. To yield transiently upon reversal, these compressive force chains must be buckled. Now, yielding involves both friction and adhesion, so that $\yield{tr}(\phi)$ increases as $\phi \to\phi_{\mu}$. 

\section{Conclusion\label{sec:conc}}

All salient aspects of our findings can be inferred from Fig.~\ref{fig:comp}. Under continuous flow, our suspensions display a steady-state yield stress, $\yield{ss}$, which emerges at $\phi_{\rm alp} \approx 0.35$, and diverges at the frictional jamming point, $\phi_{\mu} \approx 0.54$. Both critical concentrations testify to the role of friction. In contrast, oscillatory rheology reveals adhesion acting alone. The oscillatory yield stress, $\yield{os}$, diverges at the frictionless jamming point, $\phi_{\rm rcp} \approx 0.60$, and $\yield{os}(\phi) \ll \yield{ss}(\phi)$, recalling the compaction of dry grains by tapping. Finally, transient yielding under shear reversal occurs at an intermediate stress, $\yield{tr}$. It is initially close to (but slightly above) $\yield{os}$, but increases at higher $\phi$ to approach $\yield{ss}$ and diverge at $\phi_{\mu}$. 

Our results raise a number of issues for future exploration. If our microstructural proposals inspired by the notion of fragility first invoked to explain the rheology of repulsive nB suspensions are essentially correct, then a formal extension of the fragility concept to adhesive nB suspensions should prove fruitful. Figure~\ref{fig:comp} suggests that $\yield{os}$ and $\yield{ss}$ are the lower and upper bound for the yield stress of an adhesive nB suspension, because the former probes adhesion alone, while the latter probes a maximally-coupled adhesive-frictional state. Systematic investigation of other protocols besides shear reversal should test this suggestion. When combined with simulations, the results will give a more detailed understanding of the protocol-dependent yielding of such suspensions. 

Throughout, we have commented on the difference between attractive and adhesive systems. Potential attraction does not constrain interparticle motion and cannot stabilize compressive force chains. It therefore does not interact with friction in the ways that we have invoked to discuss adhesive nB suspensions. Indeed, it has been proposed~\cite{brown2010generality} and simulations suggest that yielding in attractive systems is distinct from shear jamming, with the yield stress arising from an isotropic state and simply masking shear thickening~\cite{singh2019yielding,morris2020shear}. A systematic comparison between the two kinds of suspensions remains to be done. 

Our finding that $\yield{os} \ll \yield{ss}$ suggests that the best way to lower the latter dramatically is not to perturb the interparticle adhesion (e.g. through surface `stabilizers'), but to diminish or eliminate interparticle friction. As in purely frictional nB suspensions, applying orthogonal shear or acoustic perturbations~\cite{lin2015hydrodynamic}, which mimics our oscillatory protocol, may accomplish this goal. Alternatively, one may increase $\sigma^*$ (which is $\approx 0$ in our case) until $\sigma^* > \sigma_a$. Our results suggest, and the constraint rheology model predicts, that this should lead to a drop in $\sigma_{\rm y}$ by many orders of magnitude, because yielding will no longer be dominated by friction. This insight generates a new `design principle' for surfactants as yield stress modifiers in adhesive nB suspensions, seeing these molecules as lubricants rather than stabilizers.

\section{List of Symbols}
All symbols used throughout and not defined in the Official symbols and nomenclature of The Society of Rheology are listed in Table \ref{tab:symbols}.
\noindent
\begin{table}[h!]
\renewcommand{\arraystretch}{1.2} 
    \centering
    \begin{tabular}{p{1.1cm} p{7.25cm}}
        \hline\hline
        Symbol & Definition \\
        \hline
        $a$ & Proportion of adhesive contacts in model of Guy et al.~\cite{guy2018constraint}\\
        $f$ & Proportion of frictional contacts in Wyart-Cates model~\cite{wyart2014discontinuous}\\
        $\sigma_a$ & Strength of adhesive contact\\
        $\sigma^*$ & Onset stress for frictional contacts\\
        $\phi_{\rm J}$ & Jamming volume fraction where $\eta_{\rm r} \to \infty$ as $\phi \to \phi_{\rm J}$ \\
        $\phi_{\rm rcp}$ & Random close packing, jamming point for frictionless and adhesionless particles\\
        $\phi_{\mu}$ & Jamming point for frictional but adhesionless particles\\
        $\phi_{\rm acp}$ & Adhesive close packing, jamming point for frictionless but adhesive particles \\
        $\phi_{\rm alp}$ & Adhesive loose packing, jamming point for frictional and adhesive particles\\
        $\gamma_{\rm rev}$ & Time-dependent strain upon application of a reverse stress, the opposite direction to preshear\\
        $\gamma_{\infty}$ & Long-time-limit reversal strain, $\gamma_{\rm rev}(t = \SI{D3}{\second})$, for non-flowing states\\
        $\sigma^{\prime}$ & Elastic stress in phase with applied deformation, $\sigma^{\prime} = G^{\prime}\gamma_0$\\
        $\sigma_{\max}$ & Fracture stress for parallel-plate rheometry\\
        $\sigma_y^{\rm (ss)}$ & Steady-state yield stress from minimum shear rate accessed on flow curve\\
        $\sigma_y^{\rm (os)}$ & Oscillatory yield stress defined from peak elastic stress\\
        $\sigma_y^{\rm (tr)}$ & Stress to transiently yield upon reversal, from largest increase in $\log \gamma_{\infty}$ with $\log \sigma$\\
        $\sigma_y^{\rm (pr)}$ & Stress to permanently yield upon reversal to a flowing state $\gamma_{\rm rev} \propto t$\\
        \hline \hline
    \end{tabular}
    \caption{List of symbols beyond standard rheometric symbols.}
    \label{tab:symbols}
\end{table}

\acknowledgments{This work was supported by the UK Engineering and Physical Sciences Research Council (EPSRC, EP/N025318/1, EP/L015536/1); JAR was funded by the EPSRC Centre for Doctoral Training in Soft Matter and Functional Interfaces (SOFI CDT) and AkzoNobel; and EB~by Mars Chocolate UK Ltd. Data relevant to this work can be accessed on Edinburgh DataShare at \url{https://doi.org/10.7488/ds/2634}.}

\end{document}